# Cavity Quantum Electrodynamics in the Ultrastrong Coupling Regime

**Elaheh Ahmadi[1], Hamid Reza Chalabi[2], Abbas Arab[1], and Sina Khorasani[1]**

[1] School of Electrical Engineering, Sharif University of Technology, Tehran, Iran
[2] Department of Electrical Engineering, Stanford University, Stanford, CA 94305, USA

E-mail: khorasani@sharif.edu

**Abstract.** We revisit the mathematical formulation of the famous Jaynes-Cummings-Paul Hamiltonian, which describes the interaction of a two-level atom with a single mode of an electromagnetic cavity reservoir. We rigorously show that under the condition of Ultrastrong coupling between the atom and cavity, in which the transition frequency is comparable to the coupling frequency, the bosonic field operators undergo non-sinusoidal time variations. As a result, the well-known solutions to the Jaynes-Cummings-Paul model are no longer valid, even when the rotating wave approximation is not used. We show how a correct mathematical solution could be found instead.

## 1. Introduction

The quantum mechanical description of interaction between a two-level atom and a single-mode cavity was first solved analytically by Jaynes and Cummings in 1963 [1], and independently by Paul [2]. The so-called Jaynes-Cummings-Paul (JCP) Hamiltonian was therefore shown to be exactly solvable under the Rotating-Wave Approximation (RWA) [1-3]. This field of quantum optics which studies the interaction between the quantum light emitters and cavity modes is also known nowadays as Cavity Quantum Electro-Dynamics (CQED) [4]. Hence, a simple single-mode two-level CQED system is always described in a ket space obtained by an outer product of sub-spaces corresponding to the bosonic field and the atom.

However, any such CQED system is associated with a coupling constant, which defines the strength of interaction between its sub-spaces, or partitions. Generally, a CQED system can be categorized into two distinct regimes: strong coupling and weak coupling. Modes of quantum light emitter and cavity can be in resonance in the weak coupling regime, leading to an abrupt rise in the spontaneous emission rate. Weak coupling regime is used in VCSELs [5] and LEDs [6], and more recently in generation of on-demand entangled photons [7,8]. We are not going to consider weak coupling regime in this paper and our focus is solely on the strong coupling regime.

The Strong Coupling Regime (SCR) is obtained if the coupling constant of cavity and emitter modes exceeds their decay rates. Under this condition, the cavity and emitter energy states are not degenerate anymore: they mix with each other and form new states, instead [9-11]. If the atom or the emitter is replaced by a quantum dot or quantum well, then those states could be instances of the so-called exciton-polaritons, which are the superposition of the cavity and emitter states before mixing. These new states split from each other by the amount of vacuum-rabi splitting that is proportional to coupling constant [9,11].

SCR furthermore provides an appropriate testbed for quantum information processing in solid state devices [12]. Another important feature of this regime is that cavity and emitter modes can never be in resonance with each other. This phenomenon, referred to as the anti-crossing behavior, causes anti-bunching to take place, which is in turn useful in design and realization of single photon emitters [13,14], realization of quantum encryption [15], quantum computation [10,16] and quantum repeaters [17].

The ability to fabricate high quality cavities makes it feasible for us to study the SCR in practice. SCR was first achieved between an atom and a optical cavity [18]. The equal system in solid states consists of a semiconductor micro cavity and a semiconductor quantum dot. Different teams, independently, observed SCR by exploiting different micro cavities including: micro-pillar cavities [19], photonic crystal micro-cavities [20,21] and micro-disk cavities [22].

The detuning between transition and emission frequencies can be changed by different methods. These methods include changing the lattice temperature [19,20,23], adsorption of inert gas at low temperature [24,25] and more recently, using the electrical control [26,27].

Solution of JCP model is possible in both resonant and non-resonant conditions, under which the atomic transition frequency $\omega_0$ and photon frequency $\omega_\lambda$ are respectively equal or inequal [3,4]. In a recent paper, we have devised a full mathematical solution to the most general CQED problem, encompassing an arbitrary number of emitters, each having arbitrary levels and transition rules, in coupling with a multi-mode radiation field [28], within the validity range of RWA. In order to achieve the solutions, it is customary [4] to transform the JCP Hamiltonian into the corresponding Heisenberg's interaction Hamiltonian, where the annihilation $a$ and creation $a^\dagger$ field operators are supposed to vary purely sinusoidal in time with the angular frequency of $\omega_\lambda$.

It is the purpose of this paper to mathematically show that the time-variation of field operators will no longer be sinusoidal, if the coupling strength is so large that it is comparable to the transition frequency. This occurs as a result of the fact that the transformation to Heisenberg's interaction picture needs an explicit substitution for the time dependence of field operators $a^\dagger(t)$ and $a(t)$, regardless of the RWA. These two field operators in the absence of any interaction with an emitter obey simple first-order equations of motion [4], which admit purely sinusoidal solutions having the forms $e^{\pm i\omega_\lambda t}$, which we here refer to as the free-running solution. As it will be discussed, the free-running time-dependences of $a^\dagger$ and $a$ are invalid for a sufficiently large coupling frequency, whether RWA is used, or not.

Standard configurations of nano-optical structures have reached coupling strengths $g$ as large as $400\ \mu eV$ [29], which is far below optical frequencies yet. However, extremely large coupling constants have very recently been predicted and observed at in semiconductor microcavities [30-35], magnetized two-dimensional electron gases [36], and also at microwave frequencies in cryogenic circuit quantum electrodynamics [37, known as the Ultrastrong Coupling Regime (UCR). Of course, theoretical analysis of UCR needs incorporation of anti-rotating terms; however, as we have shown here, it is the explicit time dependence of field operators which needs to be corrected as well.

We first discuss the solution with the approximation that entanglement between radiation and atom could be ignored. Then we proceed to devise the fully exact solution, taking the entanglement into account.

**2. System Hamiltonian**

In order to study the quantum interaction between the atom or the light emitter, and the single electromagnetic mode of the cavity, we model the emitter as a two-level atom and also consider the field in the cavity to be quantized, in accordance to the JCP model [1-4]

$$H = H_0 + H_1, \qquad (1)$$
$$H_0 = \hbar\omega_0 \sigma^+\sigma^- + \hbar\omega_\lambda a^\dagger a,$$
$$H_1 = -i\hbar(\sigma^+ - \sigma^-)(ga - g^*a^\dagger),$$

where $H_0$ is the basic Hamiltonian describing self-energies of the atom and the field, and $H_1$ represents the interaction terms. Also, $\omega_0$ is the transition frequency of the atom, $\omega_\lambda$ is the frequency of the mode of the cavity and, $g$ is the coupling constant between cavity mode and emission field. The operators $a$ and $a^\dagger$ are annihilation and creation of one photon, respectively, and similarly $\sigma^-$ and $\sigma^+$ are the atomic ladder operators. We have clearly ignored the zero-point energy of the radiation field, and also set the reference for the atomic energy to be zero at the mid-energy of its ground and excited state [4].

Usually, the Hamiltonian (1) is transformed to the Heisenberg's interaction picture. For a real coupling frequency $g$, this results in $H_{\text{int}} = -i\hbar g\left(\sigma^+ a e^{-i\Delta t} + \sigma^- a^\dagger e^{i\Delta t} - \sigma^- a e^{-i\Gamma t} + \sigma^+ a^\dagger e^{+i\Gamma t}\right)$ [4], where $\Delta = \omega_0 - \omega_\lambda$ and $\Gamma = \omega_0 + \omega_\lambda$ are called the detuning (difference) and sum frequencies, respectively. The transformation to the Heisenberg's interaction picture, makes the explicit use of the equations $a^\dagger(t) = e^{+i\omega_\lambda t}$ and $a(t) = e^{-i\omega_\lambda t}$. These free-running dependences are directly inserted in the transformed Hamiltonian $H_{\text{int}}$, and found from the solutions of field equations in absence of any interaction. We here show that while this is only a crude approximation for SCR, it is strongly violated in UCR. As a result, transformation to the Heisenberg's interaction picture leads to implicit equations which are more difficult to solve, than seeking the direct solution to the original Hamiltonian (1).

On one hand, when the sum frequency $\Gamma$ is much larger than the detuning $\Delta$, the RWA applies and we have $H_{\text{int}} \approx -i\hbar g\left(\sigma^+ a e^{-i\Delta t} + \sigma^- a^\dagger e^{i\Delta t}\right)$. Hence, RWA holds only if $\Gamma \gg \Delta$. On the other hand, the condition for UCR can be met if either $g\sim\Gamma$ or $g\sim\Delta$. Hence, we can easily distinguish the difference between these two cases: when the RWA is not applicable, and when we operate in UCR. Correspondingly, UCR and RWA may even be eventually compatible under special circumstances, which we do not study in this paper. Anyhow, in what follows we take care of anti-rotating terms, thus dismissing the RWA.

## 3. Ignoring Entanglement

### 3.1. Theory

In the Schrödinger's picture of motion, in contrast to Heisenberg's picture, system states are time dependent rather than operators. If we assume the system states to be spanned on the multiplication of the two states corresponding to the atom and field states, they will be entangled as time elapses as a result of light-matter interaction. For the sake of simplicity, we do consider no entanglement, and proceed to solve the Schrödinger's equations of motion. As a result, a set of wave functions are achieved, in which the atom and field states are not entangled. However, there is an analogous solution for entanglement to be taken into account which is presented in the next section. Having this said, the general eigenstate of the system is absence of entanglement takes the form

$$|\psi(t)\rangle = \left[b(t)\begin{pmatrix}1\\0\end{pmatrix} + c(t)\begin{pmatrix}0\\1\end{pmatrix}\right]|n(t)\rangle = \begin{pmatrix}b(t)\\c(t)\end{pmatrix}|n(t)\rangle, \tag{2}$$

where the state of the atom is presented as a vector and $|n(t)\rangle$ is corresponding to the number of photons. Note that $|n(t)\rangle$ is not the field's eigenket in the (2). Due to their time dependency, a set of states are achieved after solving Schrödinger's equation of motion. As it is obvious in (2), the atomic states are denoted by their equivalent vector forms. If the field operators also were defined in the matrix forms, it would be an infinite dimensions problem to solve. To relieve this complexity, we present the operators working in the space spanned by the outer product of atom and field states in the forms

$$\begin{aligned} a &\to a\begin{pmatrix}1 & 0\\0 & 1\end{pmatrix},\\ a^\dagger &\to a^\dagger\begin{pmatrix}1 & 0\\0 & 1\end{pmatrix}, \end{aligned} \tag{3}$$

$$\sigma^+ \to \sigma^\dagger \hat{1}_p,$$
$$\sigma^- \to \sigma \hat{1}_p.$$

where

$$\hat{1}_p = aa^\dagger - a^\dagger a. \tag{4}$$

is the identity operator acting in the photonic subspace. By considering these representations and using the matrix forms $\sigma^+ = \begin{pmatrix} 0 & 1 \\ 0 & 0 \end{pmatrix}$ and $\sigma^- = \begin{pmatrix} 0 & 0 \\ 1 & 0 \end{pmatrix}$, we can rewrite the Hamiltonian in (1) as

$$H = \hbar \begin{pmatrix} \omega_0 + \omega_\lambda a^\dagger a & -i(ga - g^*a^\dagger) \\ -i(ga - g^*a^\dagger) & \omega_\lambda a^\dagger a \end{pmatrix}. \tag{5}$$

The Schrödinger's equation for the operator $a$ reads

$$\frac{d}{dt}[\langle\psi(t)|a|\psi(t)\rangle] = -\frac{i}{\hbar}[\langle\psi(t)|[a,H]|\psi(t)\rangle], \tag{6}$$

where

$$[a, H] = \hbar \begin{bmatrix} \omega_\lambda a & +ig^* \\ +ig^* & \omega_\lambda \end{bmatrix}. \tag{7}$$

By substituting (2) and (7) in (6) and defining $\alpha(t) = \langle n(t)|a|n(t)\rangle$, which is the expectation value of operator $a$, it yields

$$\frac{d}{dt}\alpha(t) = -i\omega_\lambda \alpha(t) + (b(t)c^*(t) + c(t)b^*(t))g. \tag{8}$$

Using the same procedure, we find

$$\frac{d}{dt}\alpha^*(t) = +i\omega_\lambda \alpha^*(t) + \bigl(b(t)c^*(t) + c(t)b^*(t)\bigr)g^*, \tag{9}$$

where $\alpha^*(t) = \langle n(t)|a^\dagger|n(t)\rangle$.

Equations (8) and (9), as one may normally expect are complex conjugates. We now intend to solve the Schrödinger's equation for operators $\sigma^-$ and $\sigma^+$

$$\frac{d}{dt}[\langle\psi(t)|\sigma^+|\psi(t)\rangle] = -\frac{i}{\hbar}[\langle\psi(t)|[\sigma^+,H]|\psi(t)\rangle], \tag{10}$$

in which

$$[\sigma^+, H] = \hbar \begin{pmatrix} -i(ga - g^*a^\dagger) & -\omega_0 \\ 0 & +i(ga - g^*a^\dagger) \end{pmatrix}. \tag{11}$$

By substituting (2) and (11) in (10), the following ordinary differential equation is found

$$\frac{d}{dt}\big(b^*(t)c(t)\big) = \langle n(t)|(ga - g^*a^\dagger)(|c(t)|^2 - |b(t)|^2)|n(t)\rangle + i\omega_0 b(t)c^*(t) =$$
$$(|c(t)|^2 - |b(t)|^2)\big(g\langle n(t)|a|n(t)\rangle - g^*\langle n(t)|a^\dagger|n(t)\rangle\big) + i\omega_0 b^*(t)c(t). \tag{12}$$

By performing the same routine for $\sigma^-$ as we have done for $\sigma^+$, we obtain

$$\frac{d}{dt}\big(b(t)c^*(t)\big) = \langle n(t)|(ga - g^*a^\dagger)(-|c(t)|^2 + |b(t)|^2)|n(t)\rangle - i\omega_0 b(t)c^*(t) = (-|c(t)|^2 +$$
$$|b(t)|^2)\big(g\langle n(t)|a|n(t)\rangle - g^*\langle n(t)|a^\dagger|n(t)\rangle\big) - i\omega_0 b(t)c^*(t). \tag{13}$$

Obviously, and as a double-check, (12) and (13) are correctly complex conjugates. Now by defining $\beta(t) = b(t)c^*(t)$ we can rewrite (13)

$$\frac{d}{dt}\beta(t) = -i\omega_0 \beta(t) + \big(g\alpha(t) - g^*\alpha^*(t)\big)(|b(t)|^2 - |c(t)|^2). \tag{14}$$

There is a normalizing condition between $b(t)$ and $c(t)$ reading

$$|b(t)|^2 + |c(t)|^2 = 1. \tag{15}$$

At this point, and based on (15) we are able to rewrite (8) and (14) as

$$\frac{d}{dt}\alpha(t) = -i\omega_\lambda \alpha(t) + \big(\beta(t) + \beta^*(t)\big)g^* \tag{16}$$
$$\frac{d}{dt}\beta(t) = -i\omega_0 \beta(t) \pm \big(g\alpha(t) - g^*\alpha^*(t)\big)\sqrt{1 - 4|\beta|^2}. \tag{17}$$

### 3.2. Numerical results

Here, the differential equations (16) and (17) have been solved numerically. The results are illustrated in figures 1 and 2, for the case of resonance between the atom and the cavity ($\omega_0 = \omega_\lambda = 10$), or zero detuning ($\Delta = \omega_0 - \omega_\lambda = 0$). We study two cases corresponding to $g = 0.1$ and $g = 4$. Figure 1 shows the solution belonging to the $g = 0.1$, which falls in the SCR. In contrast, figure 2 shows the same for $g = 4$, which corresponds to the case of UCR.

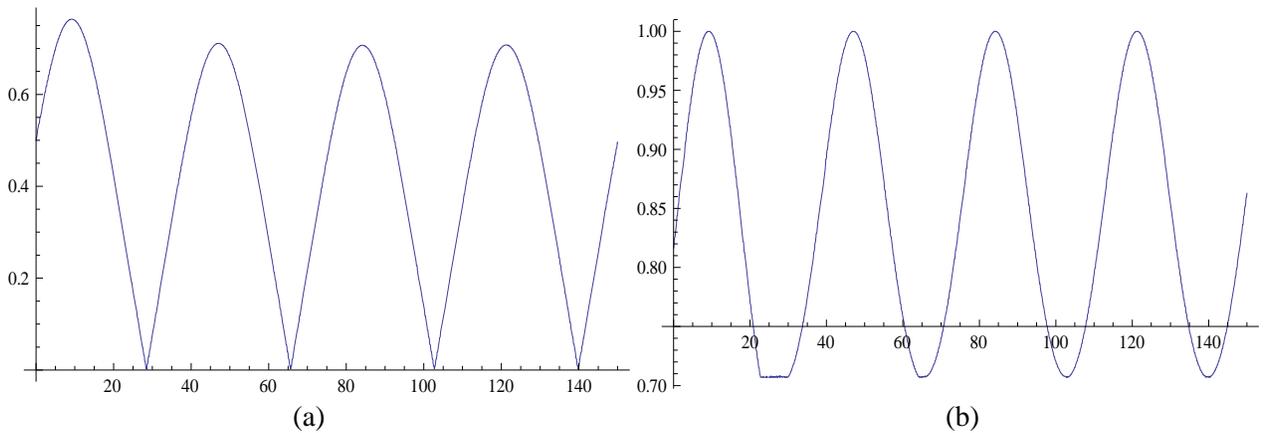

(a)  (b)

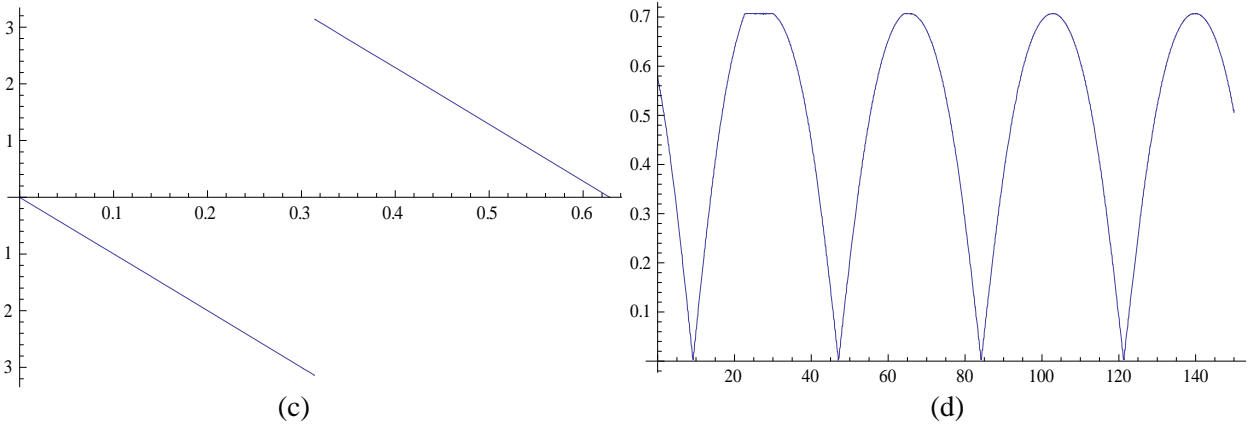

**Figure 1.** Results of integrating (16) and (17) for the resonant case with $\omega_0 = \omega_\lambda = 10$ and $g = 0.1$: (a) amplitude of $\alpha$; (b) amplitude of $c(t)$; (c) phase of $b(t)$; (d) amplitude of $b(t)$.

As it can be seen in figure 1 for SCR, all parameters belonging to the case of $g = 0.1$ vary sinusoidally, which justifies the free-running behavior of field operators. Furthermore, phase varies indistinguishably linear, which also confirms the sinusoidal time-dependence of $a^\dagger(t)$ and $a(t)$.

For comparison, we have calculated and plotted the same parameters as in figure 1 in the next figure, except that the coupling strength has been increased by a factor of 40, forcing the system to place in the UCR. This has caused the coupling frequency $g$ to become comparable to the atom-cavity frequency ($\omega_0 = \omega_\lambda = 10$), and as a result, a significant deviation from the sinusoidal free-running behavior for field operators $a^\dagger(t)$ and $a(t)$ become evident. It can be seen that the time change rate of phase even alters its sign, while in the previous case its absolute value did not vary appreciably.

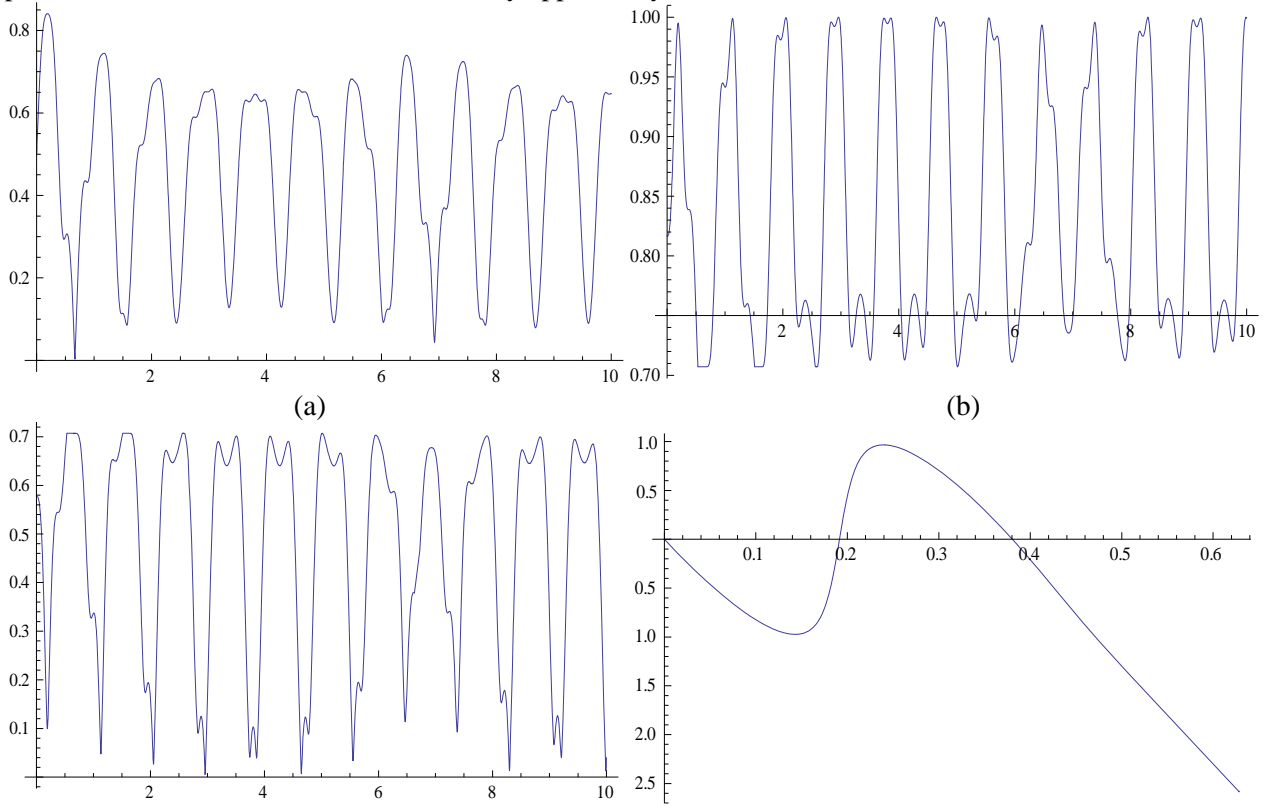

(c) (d)

**Figure 2.** Results of integrating (16) and (17) for the resonant case with $\omega_0 = \omega_\lambda = 10$ and $g = 4$: (a) amplitude of $\alpha$; (b) amplitude of $c(t)$; (c) amplitude of $b(t)$; (d) phase of $b(t)$.

We also have solved the system of equations (16) and (17) for a case that the atom and the cavity are detuned. We take $\omega_0 = 10, \omega_\lambda = 8$ with a coupling frequency of $g = 0.1$; clearly, the system is in the normal SCR. The results are depicted in figure 3.

As it can be seen here, although the amplitudes exhibit non-sinusoidal behavior, the phase keeps changing linearly with an almost constant slope. This is the signature of the validity of free-running assumption for the field operators $a^\dagger(t)$ and $a(t)$. For larger values of the coupling frequency and non-resonant case, however, we obtain comparable graphs to that of figure 2, showing strong deviation from the free-running forms.

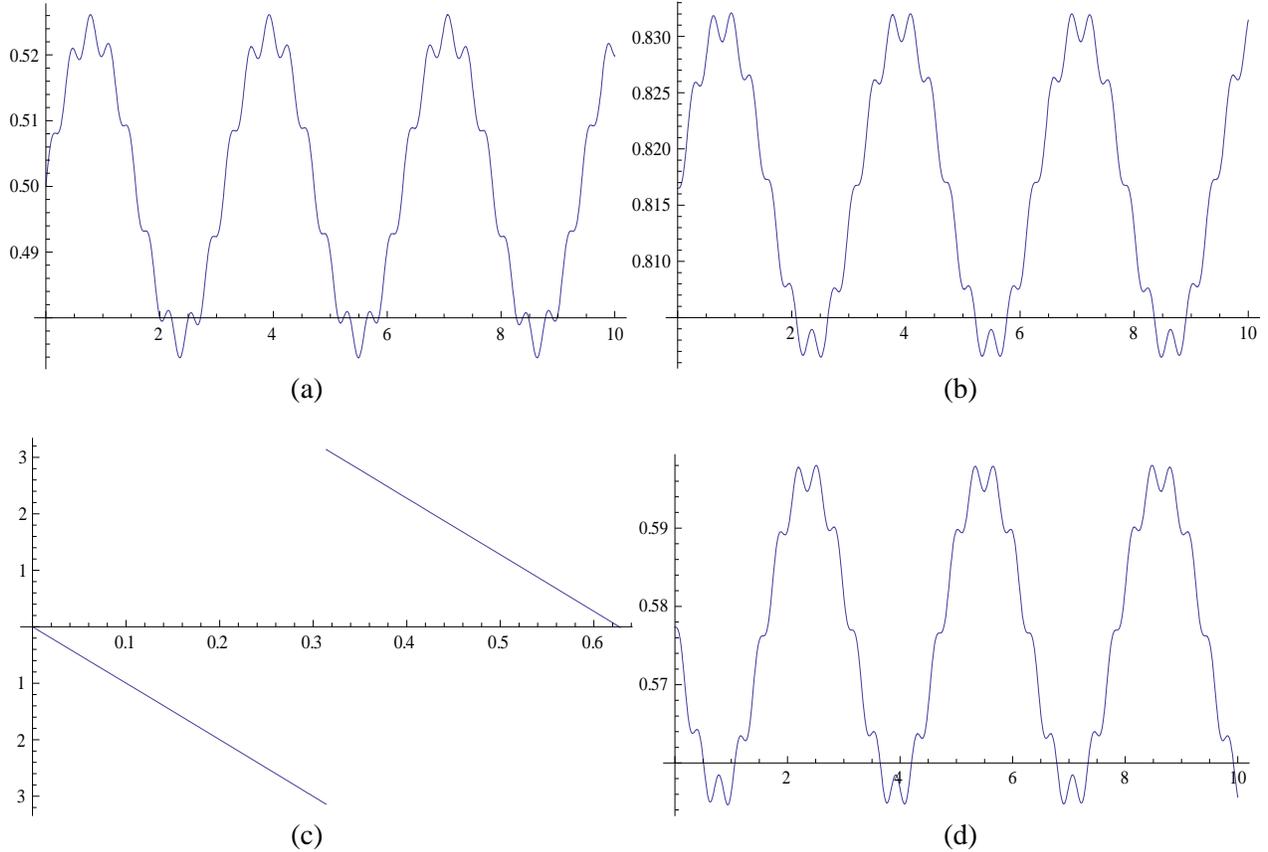

**Figure 3.** Results of integrating (16) and (17) for the non-resonant case with $\omega_0 = 10, \omega_\lambda = 8$ and $g = 0.1$: (a) amplitude of $\alpha$; (b) amplitude of $c(t)$; (c) phase of $b(t)$; (d) amplitude of $b(t)$.

## 4. Taking Entanglement into Consideration

*4.1. Theory*
In general, the state ket of an entangled atom-photon system can be written as

$$|\psi(t)\rangle = \left(\sum_{n=0} b_n(t) \begin{pmatrix} 1 \\ 0 \end{pmatrix} + \sum_{n=0} c_n(t) \begin{pmatrix} 0 \\ 1 \end{pmatrix}\right) |n\rangle = \sum_{n=0} \begin{pmatrix} b_n(t) \\ c_n(t) \end{pmatrix} |n\rangle, \tag{18}$$

where $|n\rangle$ is the eigenket of the operator corresponding to the number of the photons in the cavity and moreover, is time independent. We know from the Schrödinger's equation in quantum mechanics that:

$$\frac{d}{dt}|\psi(t)\rangle = -\frac{i}{\hbar}H|\psi(t)\rangle. \tag{19}$$

Therefore by substituting (18) into (19), we obtain

$$\frac{d}{dt}\sum_{n=0}\begin{pmatrix} b_n(t) \\ c_n(t) \end{pmatrix}|n\rangle = -i\sum_{n=0}\begin{pmatrix} \omega_0 b_n(t) + \omega_\lambda a^\dagger a b_n(t) - igac_n(t) + ig^*a^\dagger c_n(t) \\ -igb_n(t)\sqrt{n}|n-1\rangle + ig^*b_n(t)\sqrt{n+1}|n+1\rangle + n\omega_\lambda c_n(t)|n\rangle \end{pmatrix}. \tag{20}$$

We can now break (20) into two separate equations referring to each element of the vector therein. Then, we get

$$\frac{d}{dt}\sum_{n=0} b_n(t)|n\rangle =$$
$$-i\sum_{n=0}[\omega_0 b_n(t) + \omega_\lambda n b_n(t)]|n\rangle - \sum_{n=0} gc_{n+1}(t)\sqrt{n+1}\,|n\rangle + \sum_{n=1} g^* c_{n-1}(t)\sqrt{n}\,|n\rangle. \tag{21}$$

Equation (21) can also be written in the following form:

$$\frac{d}{dt}b_0(t)|0\rangle + \frac{d}{dt}\sum_{n=1} b_n(t)|n\rangle = -i\omega_0 b_0(t)|0\rangle - i\sum_{n=1}[\omega_0 b_n(t) + \omega_\lambda n b_n(t)]|n\rangle - gc_1(t)|0\rangle - \sum_{n=1} gc_{n+1}(t)\sqrt{n+1}|n\rangle + \sum_{n=1} g^* c_{n-1}(t)\sqrt{n}|n\rangle. \tag{22}$$

We separate (22) to two distinct equations, describing the unique eigenkets

$$\frac{d}{dt}b_0(t)|0\rangle = -i\omega_0 b_0(t)|0\rangle - gc_1(t)|0\rangle, \tag{23}$$

and

$$\frac{d}{dt}\sum_{n=1} b_n(t)|n\rangle =$$
$$-i\sum_{n=1}[\omega_0 b_n(t) + \omega_\lambda n b_n(t)]|n\rangle - \sum_{n=1} gc_{n+1}(t)\sqrt{n+1}|n\rangle + \sum_{n=1} g^* c_{n-1}(t)\sqrt{n}\,|n\rangle. \tag{24}$$

Following the same procedure for $c_n(t)$, we have

$$\frac{d}{dt}c_0(t) = -gb_1(t), \tag{25}$$

and

$$\frac{d}{dt}\sum_{n=1} c_n(t)|n\rangle = \sum_{n=1}(-gb_{n+1}(t)\sqrt{n+1}|n\rangle + g^*b_{n-1}(t)\sqrt{n}|n\rangle - i\omega_\lambda c_n(t)n|n\rangle). \tag{26}$$

So, we have totally four equations that can be summarized as

$$\frac{d}{dt}b_0(t) = -i\omega_0 b_0(t) - gc_1(t), \tag{27}$$

$$\frac{d}{dt}\left[b_n(t)e^{i[\omega_0+\omega_\lambda n]t}\right] = e^{i[\omega_0+\omega_\lambda n]t}\{-gc_{n+1}(t)\sqrt{n+1} + g^* c_{n-1}(t)\sqrt{n}\}, \tag{28}$$

$$\frac{d}{dt}c_0(t) = -gb_1(t), \tag{29}$$

$$\frac{d}{dt}\left[c_n(t)e^{i[\omega_\lambda n]t}\right] = e^{i[\omega_\lambda n]t}\left\{-gb_{n+1}(t)\sqrt{n+1} + g^*b_{n-1}(t)\sqrt{n}\right\}, \tag{30}$$

with the index $n$ being a positive integer. Also, $b_n(t)$ and $c_n(t)$ should satisfy the normalization condition as:

$$\sum_{n=1}(|b_n(t)|^2 + |c_n(t)|^2) = 1. \tag{31}$$

Note that this normalizing condition automatically is satisfied by the solution to the set of equations (27) through (30), and it can be used in order to check the validity of the numerical results.

### 4.2. Numerical results

We have integrated the differential equations (27) through (30), and the corresponding results are presented in the next figures. It has been demonstrated that under the condition of $\ll \omega_\lambda \approx \omega_0$, corresponding the SCR, the results converge to that of the standard JCP model, under RWA. In contrast, by violation of the above mentioned inequality, we obtained interesting results, corresponding to the UCR. Note that due to the generality of the proposed method, the initial conditions can be quite arbitrary, given that the normalization condition that must be satisfied at the initial time, too. This solution can be used in rigorous checking of the time evolution of light-matter interaction.

We first assume that system is in excited state where no photon exists. The probability of the atom to be in the excited state, under two different conditions is illustrated in figures 4 and 5. Figure 4 presents the results for the particular resonant case of $100g = \omega_\lambda = \omega_0$, which clearly falls within SCR. As it can be seen, the probability of being in the excited state varies completely sinusoidal. This again confirms the validity of the free-running assumption for the field operators under the condition $g \ll \omega_\lambda \approx \omega_0$. On the other hand, if we increase the strength of interaction and keep the system in resonance through setting $1.25g = \omega_\lambda = \omega_0$, we arrive at the solution presented in figure 5. As it can be seen, the probability of being at the excited state is no longer a sinusoidal function of time; it in fact exhibits a very complicated variation with time, which is even not quite periodic.

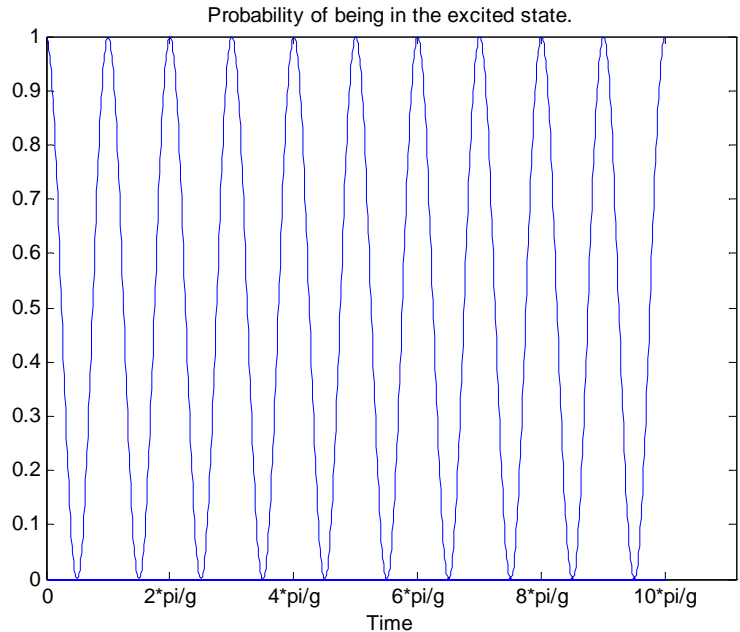

**Figure 4**. The probability of the atom to be in excited state in the case of $\omega_\lambda = \omega_0 = 100g$.

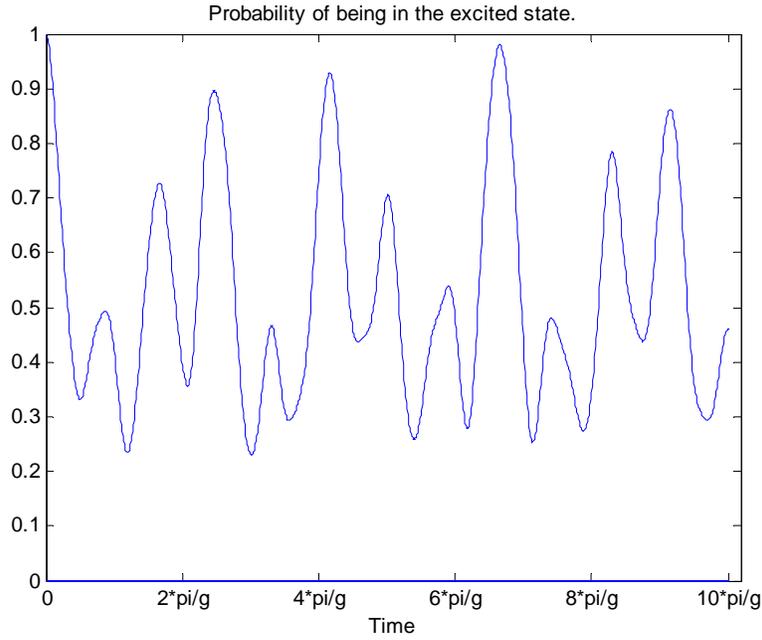

**Figure 5.** The probability of the atom to be in excited state in the case of $\omega_\lambda = \omega_0 = 1.25g$.

In addition, we have repeated the above calculations for a field state being a coherent field at time zero. A coherent state [4] is defined as $|\alpha\rangle = e^{-\frac{1}{2}|\alpha|^2} \sum_{n=0} \frac{\alpha^n}{\sqrt{n!}} |n\rangle$, where $\alpha$ is a complex number. Figures 6 and 7 present the result for $\omega_\lambda = \omega_0 = 100g$ and $\omega_\lambda = \omega_0 = 1.25g$, respectively, for comparison purposes to the last two figures. As it can be seen, deviation from the SCR results by fulfilling the assumptions of UCR is very appreciable.

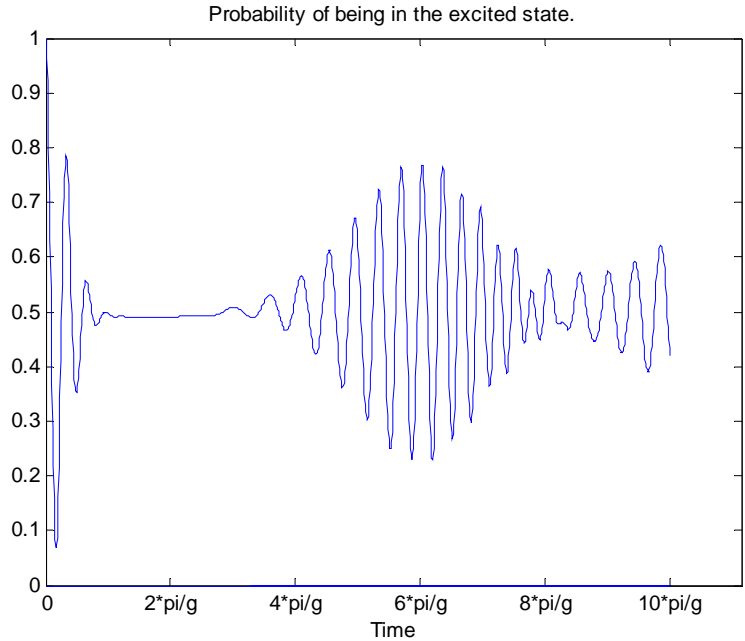

**Figure 6.** Probability for an atom to be in the excited state with the strongly interacting field starting from a resonant coherent state ($\omega_\lambda = \omega_0 = 100g$).

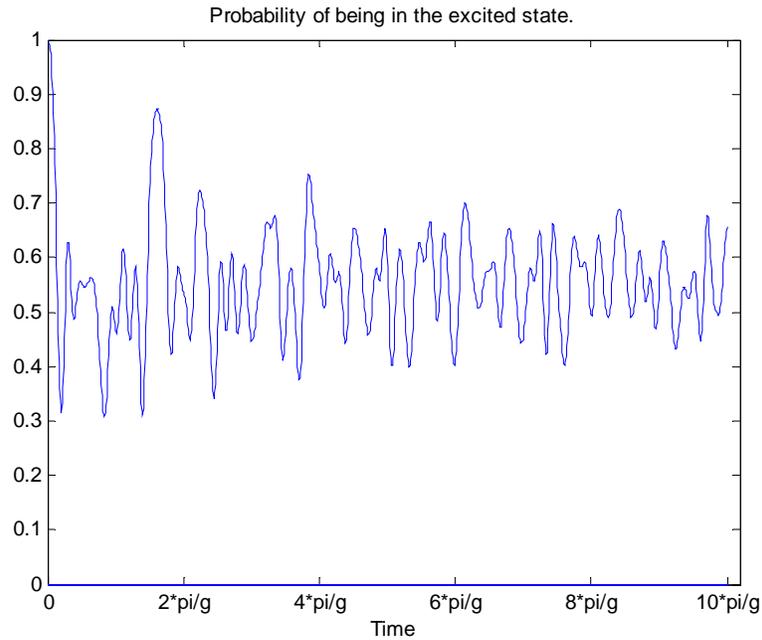

**Figure 7.** Probability for an atom to be in the excited state with the Ultrastrongly interacting field starting from a resonant coherent state ($\omega_\lambda = \omega_0 = 1.25g$).

## 5. Conclusions

We presented the direct solution to the interaction Hamiltonian of a two-level atom interacting with a single-mode electromagnetic field and arbitrary coupling strength. It has been shown that for very large values of coupling frequency, the Jaynes-Cumming-Paul solution is invalid. This happens not because of the RWA used in the standard Jaynes-Cumming-Paul formulations, but due to the non-sinusoidal temporal variations of field operators in Heisenberg's interaction picture. When the solution is sought directly in the Schrödinger's equation, it would be possible to track those non-sinusoidal behaviors of field operators, which can be quite appreciably different from the corresponding sinusoidal free-running forms.